%% file: main.tex
\documentclass[sigconf, nonacm]{acmart}
\usepackage{subfigure}
\usepackage{booktabs}
\usepackage{threeparttable}
\usepackage{algorithm, algpseudocode}
\usepackage{multirow}
\usepackage{pifont}
\usepackage{enumitem}
\newcommand\vldbdoi{XX.XX/XXX.XX}
\newcommand\vldbpages{XXX-XXX}
\newcommand\vldbvolume{14}
\newcommand\vldbissue{1}
\newcommand\vldbyear{2020}
\newcommand\vldbauthors{\authors}
\newcommand\vldbtitle{\shorttitle} 
\newcommand\vldbavailabilityurl{https://github.com/heyewuyue1/LQRS}
\newcommand\vldbpagestyle{plain} 

\begin{document}
\title{LQRS: Learned Query Re-optimization Framework for Spark SQL}

\author{
    Jiahao He, Yutao Cui, Cuiping Li, Jikang Jiang, Yuheng Hou, Hong Chen
}
\affiliation{
  Renmin University of China
}
\email{
  {hejiahao, cuiyutaooo, licuiping, jiangjikang, houyuheng, chong}@ruc.edu.cn
}






\begin{abstract}
The query optimizer is a fundamental component of database management systems. Recent studies have shown that learned query optimizers outperform traditional cost-based query optimizers. However, they fail to exploit valuable runtime observations generated during query execution to dynamically re-optimize the plan, thereby limiting further improvements in query performance. To address this issue, we propose learned query re-optimization, which allows optimization decisions to be deferred to execution time and guided by actual runtime observations. We realize this idea through LQRS, a learned query re-optimization framework that builds upon Spark SQL, exploiting runtime observations for dynamic plan refinement. Specifically, LQRS employs a curriculum reinforcement learning strategy and jointly supports pre-execution and in-execution optimization, allowing knowledge learned during execution to directly benefit pre-execution planning. Furthermore, we design a plug-and-play planner extension built upon the extensibility interfaces of Spark SQL, enabling online plan modification. Experiments on Spark SQL demonstrate that LQRS reduces end-to-end execution time by up to 90\% compared to other learned query optimizers and query re-optimization methods.
\end{abstract}

\maketitle

\pagestyle{\vldbpagestyle}
\begingroup\small\noindent\raggedright\textbf{PVLDB Reference Format:}\\
\vldbauthors. \vldbtitle. PVLDB, \vldbvolume(\vldbissue): \vldbpages, \vldbyear.\\
\href{https://doi.org/\vldbdoi}{doi:\vldbdoi}
\endgroup
\begingroup
\renewcommand\thefootnote{}\footnote{\noindent
This work is licensed under the Creative Commons BY-NC-ND 4.0 International License. Visit \url{https://creativecommons.org/licenses/by-nc-nd/4.0/} to view a copy of this license. For any use beyond those covered by this license, obtain permission by emailing \href{mailto:info@vldb.org}{info@vldb.org}. Copyright is held by the owner/author(s). Publication rights licensed to the VLDB Endowment. \\
\raggedright Proceedings of the VLDB Endowment, Vol. \vldbvolume, No. \vldbissue\ %
ISSN 2150-8097. \\
\href{https://doi.org/\vldbdoi}{doi:\vldbdoi} \\
}\addtocounter{footnote}{-1}\endgroup

\ifdefempty{\vldbavailabilityurl}{}{
\vspace{.3cm}
\begingroup\small\noindent\raggedright\textbf{PVLDB Artifact Availability:}\\
The source code, data, and/or other artifacts have been made available at \url{\vldbavailabilityurl}.
\endgroup
}

\input{Introduction}
\input{RelatedWork}
\input{SystemOverview}
\input{DecisionModel}
\input{PlannerExtension}
\input{Experiments}
\input{Conclusion}

\clearpage

\bibliographystyle{ACM-Reference-Format}
\bibliography{main}

\end{document}

%% file: Introduction.tex
\section{Introduction}

The query optimizer is a fundamental component of modern database management systems. Its primary role is to select an efficient execution plan for each query. Traditional cost-based query optimization~\cite{traddp} is a common technique in query optimizers, where the idea is to select the plan with the minimum estimated cost. However, these cost models employ a variety of handcrafted formulas to approximate actual execution costs; such approximations are often inaccurate, resulting in suboptimal execution plans. 

To address this problem, people propose to employ machine learning in query optimization~\cite{marcus2021bao, anneser2023autosteer, fastgres, scope,chen2024lero, zhong2024foss, deepo, hybrid, leon, neo, balsa, loger, glo}. Such a learned query optimizer (LQO) leverages models trained on historical execution feedback to improve the enumeration and evaluation of candidate execution plans, thereby enabling more effective optimization decisions.

Despite their effectiveness, most learned query optimizers such as Lero~\cite{chen2024lero} often suffer from limitations imposed by the optimize-then-execute paradigm: they optimize queries based on estimated statistics and then execute a static query plan (referred to as \textbf{pre-execution} optimization). As a result, they fail to exploit valuable runtime observations generated during query execution to dynamically re-optimize the plan, thereby limiting further improvements in query performance. For example, as shown in Table~\ref{tab:mot}, STACK Q1\#1 involves optimizing a join plan over four tables \textbf{\texttt{t}}, \textbf{\texttt{s}}, \textbf{\texttt{q}}, and \textbf{\texttt{tq}}. Lero selects the join plan (\textbf{\texttt{t}}$\bowtie$\textbf{\texttt{tq}})$\bowtie$(\textbf{\texttt{s}}$\bowtie$\textbf{\texttt{q}}) based on its learned model, resulting in an execution time of 21.52s. This plan is fixed and cannot be modified during execution. However, runtime observations reveal that scanning table \textbf{\texttt{t}} produces only a single tuple. Leveraging this feedback, the join plan can be re-optimized (referred to as \textbf{in-execution} optimization) to \textbf{\texttt{t}}$\bowtie$\textbf{\texttt{tq}}$\bowtie$\textbf{\texttt{q}}$\bowtie$\textbf{\texttt{s}}, achieving a significantly shorter execution time of 15.24s. 

\begin{table}[t]
  \caption{STACK Q1\#1 involves optimizing a join plan over $t$, $s$, $q$, and $tq$. Lero fixes the join order before execution, whereas LQRS adjusts the join order based on runtime information, resulting in a shorter execution time.}
  \label{tab:mot}
  \begin{threeparttable}
      \begin{tabular}{ccc}
        \toprule
        Plan & Lero~\cite{chen2024lero} & LQRS (ours)\\
        \midrule
        Pre-execution & $(t\bowtie tq)\bowtie (s\bowtie q)$ & $q\bowtie tq\bowtie t\bowtie s$ \\
        In-execution  & Fixed during execution & $t\bowtie tq\bowtie q\bowtie s$\tnote{*}\\
        Execution time & 21.52s & \textbf{15.24s}\\
        \bottomrule
      \end{tabular}
      \begin{tablenotes}
        \footnotesize
        \item[*] During execution, LQRS observes that the scan on table $t$ produces only one tuple and reorders the join accordingly.
      \end{tablenotes}
  \end{threeparttable}
\end{table}

Such a limitation goes beyond LQO's expressiveness or feature engineering. For example, characteristics such as the value distribution of aggregation keys or the exact selectivity of complex predicates only manifest once data is actually processed~\cite{onlineagg}. Even with sophisticated encodings and learned estimators, LQO models restricted to \textbf{pre-execution} optimization frequently struggle to produce reliable predictions for such properties, and empirical studies have shown that learned estimators often fail to consistently outperform traditional cost-based optimizers such as PostgreSQL’s planner~\cite{heinrich2025good}. This fact suggests that the challenge arises not merely from modeling limitations, but from making decisions without access to critical runtime observations. This motivated us to think about an important question: \textbf{how can we design a learned query re-optimization framework that can collect runtime observations and use them to dynamically re-optimize previously generated query plans during query execution?} 

\begin{figure}[t]
\centerline{\includegraphics[width=\linewidth]{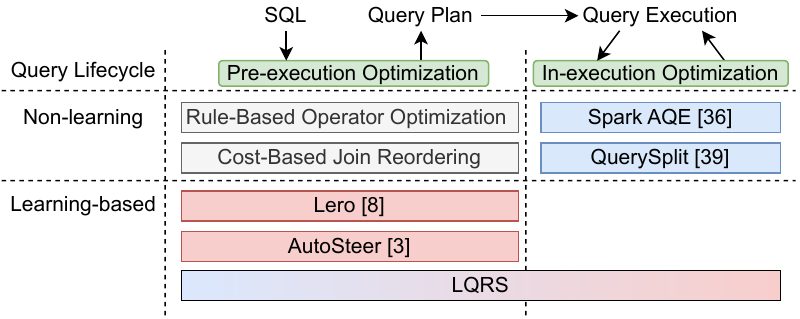}}
\caption{LQRS is the first learned query optimizer achieving both pre-execution and in-execution query plan optimization on Spark SQL.}
\label{fig:methodology}
\end{figure}

To address this issue, we propose learned query re-optimization, which allows optimization decisions to be deferred to execution time and guided
by actual runtime observations. We realize this idea through LQRS, a \underline{l}earned \underline{q}uery \underline{r}e-optimization framework that exploits runtime observations for dynamic plan refinement. Our realization of LQRS builds upon the Adaptive Query Execution (AQE) foundation of \underline{S}park SQL~\cite{sparkaqe}, which provides the necessary runtime hooks for collecting execution feedback and revising optimization decisions. As illustrated in Figure~\ref{fig:methodology}, existing query optimization techniques intervene at different phases of a query’s lifecycle: traditional rule-based operator optimization and cost-based join reordering are performed entirely before execution; most learned query optimizers such as Lero\cite{chen2024lero} and AutoSteer~\cite{anneser2023autosteer} are also restricted to the pre-execution phase; while adaptive query processing techniques like Spark AQE~\cite{sparkaqe} and QuerySplit~\cite{querysplit} adjust execution without leveraging learning-based guidance. \textbf{In contrast, LQRS represents the first learned query optimizer on Spark SQL that jointly supports pre-execution and in-execution plan refinement, effectively unifying learned query optimization with adaptive query processing.}

LQRS continuously collects runtime execution statistics and exposes them as plan features, including actual cardinality observed during query execution. These runtime features allow the optimizer to analyze partially executed plans using information unavailable before execution, enabling more effective optimization decisions during query processing. To enhance learning stability, LQRS adopts a curriculum learning strategy: it starts with simpler decisions and gradually increases the complexity, enabling the model to learn more effective strategies in a step-by-step manner. In addition, LQRS unifies pre-execution and in-execution plan optimization within a single learning process, allowing in-execution learned knowledge to be leveraged for pre-execution optimization.

To support execution-time learning and re-optimizing, LQRS extends the Spark AQE framework to provide fine-grained runtime feedback and intervention capabilities. The planner extension exposes query stage-level execution signals like changes in shuffle counts and the structure of partially executed plans during execution, allowing LQRS to receive dense feedback beyond end-to-end query latency. This design produces feedback that is at least 3$\times$ denser than traditional end-to-end learning approaches. Additionally, the planner extension allows LQRS to directly intervene in partially executed plans, enabling dynamic re-optimization such as bushy join exploration during execution. 

Our contributions can be summarized as follows:
\begin{itemize}[nosep, leftmargin=*]
    \item We design and implement \emph{LQRS}, a learned query re-optimizer which incorporates reinforcement learning into the Spark AQE framework to make decisions during query execution based on actual runtime observations.

    \item We adopt a decision model built upon the TreeCNN architecture ~\cite{tcnn}, which learns optimization actions directly from query stage-level feedback collected during query execution. By using true cardinalities and size in bytes as input features, the model achieves better query performance with a simplified architecture. 
    
    \item We design a plug-and-play planner extension built upon the extensibility interfaces of Spark SQL, enabling online plan modification during execution. The extension intercepts runtime execution plans, interacts with the decision model, and applies optimization actions such as join reordering.
    
    \item We conduct extensive experiments on four public benchmarks. Experiments on Spark SQL demonstrate that LQRS reduces end-to-end execution time by up to 90\% compared to other learned query optimizers and query re-optimization methods.
\end{itemize}

%% file: RelatedWork.tex
\section{Related Work}

\begin{figure*}[t]
\centerline{\includegraphics[width=\linewidth]{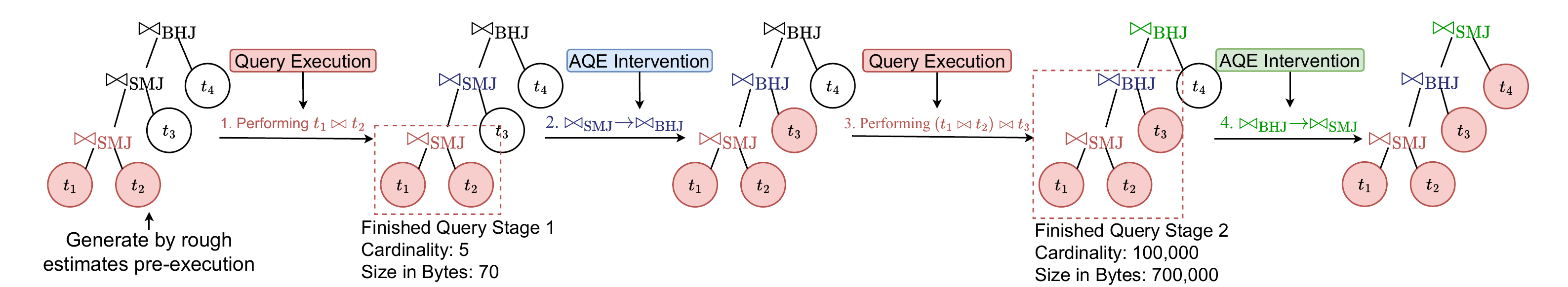}}
\caption{This query performs three joins: after the first join $t1 \bowtie t2$ produces only 5 rows, AQE switches the second join from SMJ to BHJ; after the second join $(t1 \bowtie t2) \bowtie t3$ produces 100,000 rows, AQE switches the third join from BHJ to SMJ.}
\label{fig:aqe}
\end{figure*}

\paragraph{Learned Query Optimization} 

Since LEO~\cite{leo} in 2001 introduced the concept of a learned query optimizer, recent studies on LQO have evolved into two main directions: plan-constructors and plan-steerers. Plan-constructor methods aim to build query plans directly from scratch. These approaches enable the discovery of novel and potentially more efficient plans. 

Neo ~\cite{neo} pioneers this category by leveraging past queries to progressively construct improved query plans, using the native optimizer as a weak oracle. Balsa ~\cite{balsa} further improves the convergence efficiency of plan learning by employing enhanced initialization and training techniques. LOGER ~\cite{loger} enhances interpretability in learned planning by detecting detrimental operators and adjusting reward weighting accordingly. GLO ~\cite{glo} improves generalization to unseen workloads by clustering database statistics and learning generalized policies across these clusters. 

Plan-steerer methods improve query plans by exploiting the internal mechanisms of the native optimizer. Instead of constructing plans from scratch, they guide the optimizer to explore multiple candidate plans and select the one with the lowest estimated cost. For example, Bao ~\cite{marcus2021bao}, DeepO ~\cite{deepo}, SCOPE ~\cite{scope}, FASTgres ~\cite{fastgres}, and AutoSteer ~\cite{anneser2023autosteer} generate different query plans by toggling rule-based heuristics (i.e., hint sets) within traditional optimizers. HybridQO ~\cite{hybrid} explores alternative join orders by injecting leading hints. Lero~\cite{chen2024lero} modifies cardinality estimates in the native optimizer to produce diverse plans.

LQRS differs from existing LQOs in that it leverages runtime plan feature representations to make more informed optimization decisions and allows optimization decisions made pre-execution to be corrected during query execution.

\paragraph{Adaptive Query Processing}

Adaptive query processing (AQP) has been extensively investigated to facilitate query plan adjustments during execution based on runtime feedback ~\cite{scramble, smooth, stairs, aqp, mid, ajp, lip, justen2024polar, udao}. The primary AQP techniques are categorized as operator-based AQP and plan-based AQP.

Operator-based AQP adapts the behavior of physical operators during execution. For instance, Lookahead Information Passing~\cite{lip} generalizes semi-join techniques to improve the pipelining of equijoins. The Adaptive Join Algorithm~\cite{ward2019sql} dynamically selects between hash join and nested loop join during runtime.

Plan-based AQP (or query re-optimization) re-invokes the optimizer at certain points during query execution to revise structural decisions such as join orders or the execution order of subqueries ~\cite{dynamicreopt,pop,incollection, hadoop,arqo,lpce,querysplit}. Following the pioneering work on dynamic re-optimization~\cite{dynamicreopt}, Adaptive Metaprogramming~\cite{arqo} moves recursive query optimization and code generation from compile time to runtime, enabling customized join-order optimization based on runtime information. LPCE~\cite{lpce} adopts a two-stage learning strategy, consisting of an initial model and a refinement model, to progressively improve cardinality estimation accuracy for query operators. QuerySplit~\cite{querysplit} proposes a proactive query re-optimization framework that decomposes a query into multiple logical subqueries and determines their execution order according to estimated costs and output cardinalities.

While most AQP techniques are not learning-based, LQRS extends the plan-based AQP paradigm and leverages reinforcement learning to guide optimization decisions based on runtime feedback. It also integrates pre-execution and in-execution optimization into a unified learning process and supports adaptation at both the operator and plan levels.

%% file: SystemOverview.tex
\section{Illustrative Overview}

\subsection{Spark SQL’s Query Execution Model}
In Spark SQL, query execution is organized around query stages, which serve as the fundamental units of physical execution and runtime optimization. A \textbf{query stage} corresponds to a contiguous subgraph of the physical plan that can be executed without shuffle boundaries; stages are separated by shuffle operations that materialize intermediate results (for example, when a join is performed on a different join key from the preceding join, a shuffle is required to repartition the data, thereby introducing a new query stage). Once a query stage completes, Spark obtains accurate runtime statistics for that stage, including output cardinalities and data sizes, which are unavailable prior to execution.

\subsection{Spark SQL’s Adaptive Query Execution}
In order to tackle the challenges of cardinality estimation and physical plan cost estimation, Spark SQL 3.0 (2020) introduced AQE. AQE starts with an initial pre-execution query plan based on rough estimations and then re-optimizes it during execution using actual runtime observations. 

Figure~\ref{fig:aqe} illustrates a mechanism adopted by AQE called \emph{dynamic join selection}. In Figure~\ref{fig:aqe}, the query is executed in multiple query stages. The red segments in the figure correspond to standard query execution without adaptation, while the blue and green parts indicate the two plan modifications introduced by AQE interventions at runtime. After the first query stage, which executes the join $t1 \bowtie t2$ and produces only 5 rows and 70 bytes of intermediate result, AQE decides to switch the join strategy for the second stage from SMJ to BHJ. After the second stage, which executes $(t1 \bowtie t2) \bowtie t3$ and produces 100,000 rows and 700,000 bytes of intermediate result, AQE again adapts the plan, switching the join strategy for the third stage from BHJ back to SMJ. By reasoning at the granularity of query stages during execution, AQE can make more informed decisions based on actual intermediate cardinalities rather than relying solely on rough pre-execution estimates.

\subsection{Spark AQE's Limitations}

Despite its ability to adapt join strategies at runtime, Spark AQE has inherent limitations. First, the effectiveness of AQE’s dynamic join selection is constrained by human-configured thresholds. In particular, AQE only converts an SMJ to a BHJ if the size of one side's intermediate result is smaller than the broadcast join threshold (BJT). Since broadcasting a large table can lead to worker crashes, administrators typically configure BJT with an over-conservative static value, which limits the performance benefits.

Second, AQE cannot change the join order once query execution begins. The optimizer relies on the pre-execution plan to determine the sequence of joins. As a result, queries with suboptimal join orders may still incur significant performance penalties, leaving untapped optimization potential that current AQE cannot address.

\begin{figure*}[t]
\centerline{\includegraphics[width=\linewidth]{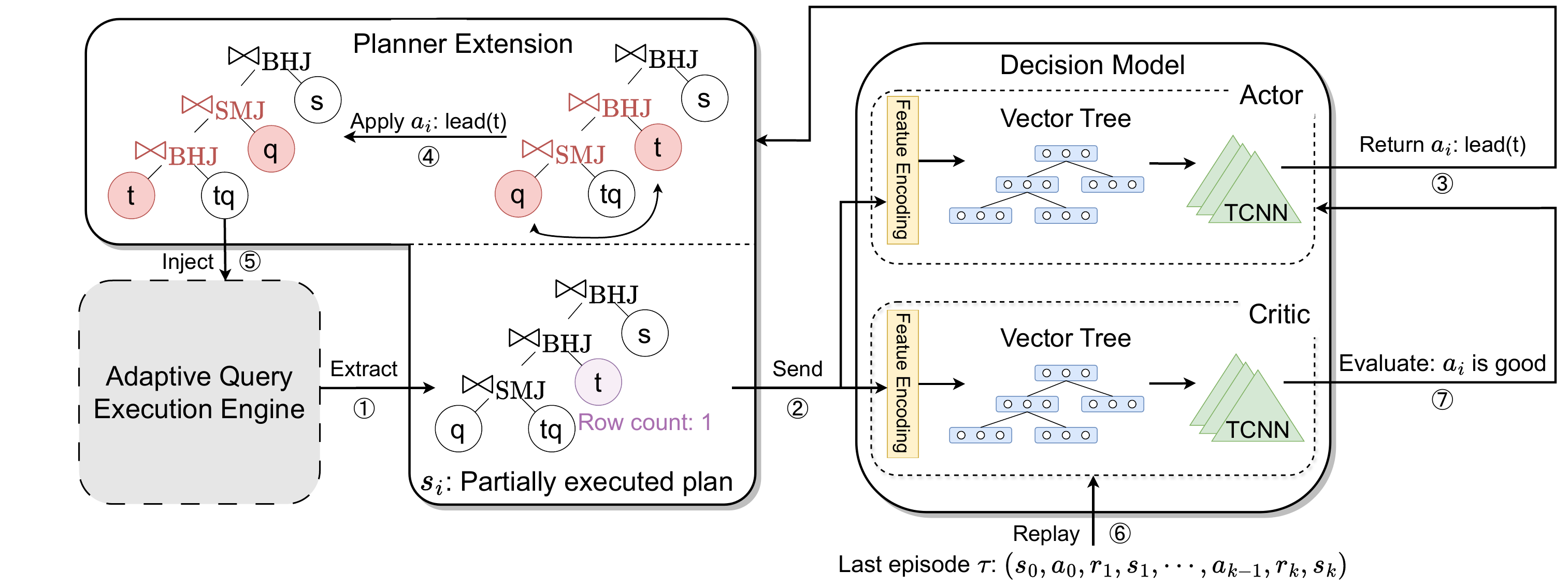}}
\caption{An overview of LQRS, which consists of two main components: (1) a decision model that generates optimization actions, and (2) an AQE planner extension that applies these actions and provides feedback to the decision model.}
\label{fig:LQRS}
\end{figure*}

\subsection{Our Solution}

LQRS is tightly integrated with Spark SQL’s query execution model. It treats the completion of a query stage as a natural decision point, where partially executed plans and query stage-level statistics are exposed to the decision model. By aligning optimization actions with query stage boundaries, LQRS ensures that runtime plan adaptations are both semantically valid and practically efficient within Spark SQL’s execution framework.

Next, we introduce the main components of LQRS and its basic workflow. As shown in Figure~\ref{fig:LQRS}, LQRS consists of two core components: a \emph{planner extension} and a \emph{decision model}. The figure illustrates an example where LQRS performs a single optimization action during the execution of the STACK query q1\#1.

\begin{enumerate}[nosep, leftmargin=*]
    \item When Spark SQL's AQE is triggered during execution, the planner extension extracts (\ding{172}) the current query plan along with actual cardinality statistics and sends (\ding{173}) it to the decision model. In Figure~\ref{fig:LQRS}, the system has just completed scanning table \textbf{\texttt{t}} and observed the output cardinality. The planner extension then sends the partially executed plan and the row count (1) to the decision model. This process can also be triggered during the query optimization phase, in which case actual cardinalities are not available.
    
    \item The decision model analyzes the input plan and returns (\ding{174}) optimization actions. As shown in Figure~\ref{fig:LQRS}, the decision model observes that table \textbf{\texttt{t}} produces only one output row but is joined after two larger tables. The actor therefore infers that placing \textbf{\texttt{t}} earlier in the join order could be more efficient and outputs the action lead(\textbf{\texttt{t}}).
    
    \item The planner extension applies (\ding{175}) these actions to the ongoing query plan and injects (\ding{176}) the optimized plan back to the AQE engine to resume execution. As illustrated in Figure~\ref{fig:LQRS}, after receiving the action lead(\textbf{\texttt{t}}), the planner extension attempts to reposition table \textbf{\texttt{t}} lower in the plan, allowing it to participate in the join operation earlier. In coordination with the planner extension's reposition behavior, the AQE engine can further adapt the join strategies, transforming the original SMJ–BHJ–BHJ sequence into a more efficient BHJ–SMJ–BHJ sequence.
     
     \item After the query completes after $k$ iterations of \ding{172}-\ding{176}, the decision model replays (\ding{177}) the collected $k$ step optimization process to learn from past decisions and improve its optimization strategy for future queries. During this replay process, the critic evaluates (\ding{178}) each action taken by the actor and provides feedback signals, which are used to guide policy improvement.
\end{enumerate}

We formalize the optimization process of LQRS as an \textit{episode}: 
\begin{displaymath}
    \tau=(s_0, a_0, r_1, s_1,\cdots,a_{n-1},r_k,s_k)
\end{displaymath}
where each state $s_i$ consists of a partially executed plan $p_i$ and its corresponding statistics $c_i$, while $a_i$ denotes the action taken at step $i$, and $r_{i+1}$ represents the shaping reward obtained after taking action $a_i$. Details will be given in the Section \ref{sec:sarsa}.

\subsection{Our Result}

As a result, compared to Spark AQE, LQRS reduces the runtime from 33.53s to 15.24s in the above example, achieving over 50\% lower execution latency. In this example, Spark AQE is unable to reorder joins based on runtime observations and thus continues to execute the query using a suboptimal join order determined at pre-execution time, which both incurs higher execution cost and generates large intermediate results, preventing effective use of dynamic join selection. In contrast, LQRS captures runtime signals that reveal a better join order and proactively performs join reordering during execution, after which the AQE engine replans the remaining query with a more efficient join strategy.

%% file: DecisionModel.tex
\section{Decision Model}

\subsection{The Actor-Critic Framework}

We adopt the actor–critic framework to fully exploit the frequent feedback signals from runtime, which promotes faster convergence and allows us to observe the critic’s reward or penalty for each action taken by the actor.

\subsubsection{State, Action, and Reward}
\label{sec:sarsa}

To introduce the framework, we formally define the three main components of the Markov decision process for query optimization:

\noindent\textbf{State.} Each state $s_t=(p_t, c_t)$ consists of:
\begin{itemize}[nosep, leftmargin=*]
    \item $p_t$: the partially executed plan at time step $t$;
    \item $c_t$: the true cardinalities and output size in bytes of completed query stages, obtained from AQE statistics.
\end{itemize}

\noindent\textbf{Action Space.} LQRS supports five actions: controlling the plan initialization strategy, swapping the $i$-th and $j$-th leaf node, joining the leaf nodes $t_1,t_2,\cdots,t_l$ first if possible, setting broadcast join hints on leaf nodes to enforce broadcast joins, and intentionally taking no action for now. The notations and definitions of these actions are summarized in Table~\ref{tab:actions}.

\begin{table}[t]
\caption{Action space of LQRS's decision model}
\begin{center}
\begin{tabular}{cc}
    \toprule
    Notation & Definition\\
    \midrule
    init$(1/0)$ & select plan initialization strategy\\
    swap$(i,j)$ & swap the $i$-th and $j$-th leaf node\\
    lead$(t_1,t_2,\cdots,t_l)$ & join $t_1,t_2,\cdots,t_l$ first if possible \\
    broadcast$(i)$ & broadcast the $i$-th leaf node\\
    no-op$()$ & take no action for now\\
    \bottomrule
\end{tabular}
\label{tab:actions}
\end{center}
\end{table}

\begin{itemize}[nosep, leftmargin=*]
    \item The \textbf{\textit{init}} action controls the initial plan construction strategy, determining how the starting execution plan is generated before any runtime adaptation takes place. In some cases, a traditional cost-based initialization produces a plan that is already close to optimal, while in others, a rule-based initialization offers a better starting point for subsequent optimization actions. Allowing the decision model to dynamically select between different plan initialization strategies helps LQRS to escape local optima and discover better execution plans.
    \item The \textbf{\textit{swap}} action enables fine-grained join reordering by exchanging two leaf nodes in the current logical plan. This action provides local flexibility for adjusting join order and is particularly effective when small structural changes are sufficient to improve performance. 
    \item The \textbf{\textit{lead}} action promotes a specific table, or a subset of tables, to be joined earlier in the execution plan. Compared to the \textit{swap} action, this action offers a more direct and efficient mechanism for correcting suboptimal join orders, as many beneficial transformations can be achieved by repositioning a single critical leaf node. As a result, the \textit{lead} action serves as a practical control for guiding join order decisions.
    \item The \textbf{\textit{broadcast}} action enforces a broadcast join on a selected leaf node. This action is designed to overcome the overly conservative behavior induced by static broadcast thresholds. Although forcing a broadcast join introduces some potential risks, it enables the correction of conservative join method choices.
    \item The \textbf{\textit{no-op}} action allows the decision model to intentionally defer optimization decisions and wait for additional runtime information. This action is particularly useful in early execution stages where limited feedback is available and premature interventions may be harmful. By supporting deliberate inaction, LQRS can balance exploration and stability while reducing unnecessary plan perturbations.
\end{itemize}

Together, these actions enable LQRS to perform both coarse-grained and fine-grained plan adaptations across different stages of query execution.

\noindent\textbf{Reward.} 
A key challenge of runtime plan adaptation is that modifying an execution plan after some query stages have already materialized may require redistributing intermediate data. In Spark, this typically manifests as additional shuffle operations. For example, consider a query whose initial execution plan decides to evaluate $t_1 \bowtie t_2$ first, followed by a join with $t_3$. During execution, Spark scans $t_1$ and $t_2$ and performs a shuffle to partition their records according to the join key of $t_1 \bowtie t_2$, preparing for a shuffle-based join. If a subsequent runtime decision swaps the join order to instead evaluate $t_1 \bowtie t_3$ first, the existing partitioning of $t_1$, which was produced to satisfy the join with $t_2$, is no longer compatible with the new join requirement. As a result, $t_1$ is reshuffled to match the join key of $t_1 \bowtie t_3$, allowing execution to proceed correctly under the updated plan.

Motivated by this observation, LQRS explicitly incorporates reshuffles as intermediate feedback in the reward definition, encouraging the agent to account for the operational cost of reorganizing intermediate data when making runtime decisions. Based on empirical validation, we define the total reward as:
\begin{displaymath}
    R(\tau)=\sum_{i=1}^k \gamma^{i-1}r_i-\sqrt{T_\text{execute}(\tau)} ,
\end{displaymath}
where $T_\text{execute}(\tau)$ denotes the total execution time of episode $\tau$ (in seconds). The square root operation is used to reduce reward scale variations across queries. If execution fails due to OOM errors or exceeds the time limit of 300s, $R(\tau)$ is assigned a substantial negative value (e.g., $-\sqrt{300}$).

The intermediate reward $r_i$ is defined as the negative number of additional shuffle operations introduced after taking action $a_i$, normalized by 10. This normalization aligns the scale of intermediate rewards with that of $\sqrt{T_\text{execute}(\tau)}$ under our experimental settings, ensuring that intermediate and terminal rewards are of comparable magnitude during training. This formulation naturally provides positive feedback when the agent selects a \textit{no-op} action (i.e., no additional shuffle is introduced), encouraging conservative behavior in early training, while penalizing actions that trigger reshuffles and discouraging costly plan changes at later stages. The discount factor $\gamma < 1$ further ensures that accumulated intermediate penalties do not dominate the overall reward.

\subsubsection{Policy and State Transition}

The actor network implements a stochastic policy $\pi_\theta(a|s)$, which maps states to action probabilities through softmax normalization:

\begin{displaymath}
    \pi_\theta(a|s) =
    \begin{cases}
        P_1, \quad &\text{if} \ a=a_1 \ (\text{e.g.,swap}(1,2))\\
        P_2, \quad &\text{if} \ a=a_2 \ (\text{e.g.,broadcast}(1))\\
        &\vdots \\
        P_m, \quad &\text{if} \ a=a_m \ (\text{e.g., no-op}())
    \end{cases}
\end{displaymath}
where $\sum_{i=1}^m P_i=1$. The probabilities $P_i$ are computed by the actor network with parameters $\theta$ via softmax normalization over its output logits. This probabilistic approach enables exploring alternative optimization paths while maintaining likely good choices. 

After the actor takes an action $a$, the current state $s$ transitions to the next state $s'$. The \textit{state transition} probability $P(s'|s,a)$ reflects the uncertainty in adaptive execution:

\begin{displaymath} 
    P(s'|s, a) =
    \begin{cases}
        P_1, \quad &\text{if} \ s'=s_1 \ (\text{immediate AQE trigger}) \\
        P_2, \quad &\text{if} \ s'=s_2 \ (\text{delayed 1 stage})  \\
        &\vdots \\
        P_h, \quad &\text{if} \ s'=s_h \ (\text{delayed multiple stages})  \\
    \end{cases}
\end{displaymath}
The uncertainty arises from Spark AQE's non-deterministic triggering: the execution engine may complete a varying number of stages between optimizations. As a result, applying the same action $a$ in state $s$ can lead to different successor states $s'$.

\subsubsection{State Value and Action Value}

The \textit{state value} $v_\pi(s)$ quantifies the expected long-term utility of being in state $s$. The critic evaluates this value based on the current partial plan and runtime statistics. For example, if a join operation involves high cardinalities on both sides—often indicating expensive data shuffling—the critic may assign a lower $v_\pi(s)$ compared to states with smaller intermediate results. Note that this pattern isn't universal—while some large joins may be unavoidable for optimal performance, it serves as an intuitive starting point. Through training, the critic learns to refine these initial heuristics into precise value estimates.

Also notice that the state value $v_\pi(s)$ depends not only on the current plan $s$ but also on the actor's policy $\pi$. Even when reaching a promising state $s$, a suboptimal policy may select actions that degrade subsequent states. For example, a policy favoring broadcast joins for large tables could turn $s$ (with moderate cardinalities) into $s'$ (with memory pressure), lowering $v_\pi(s)$ despite $s'$'s inherent potential. The critic's estimated value $v_\phi(s)$ approximates $v_\pi(s)$, learning to anticipate these policy-induced outcomes through exposure to the actor's decision patterns.

The \textit{action value} $q(s_t,a_t)$ measures the impact of action $a_t$ on optimization potential. Let $v_\phi(s)$ denote the critic's value estimate with parameters $\phi$. We compute:
\begin{displaymath}
    q(s_t, a_t) = r_{t+1}+v_\phi(s_{t+1})-v_\phi(s_{t})
\end{displaymath}
where $s_{t+1}$ is the state after performing $a_t$. For instance, swapping larger tables early (action $a_t$) might decrease $v_\phi(s_{t+1})$ due to increased shuffle costs, yielding negative $q(s_t,a_t)$. Conversely, broadcasting a small table (action $a_t'$) could improve $v_\phi(s_{t+1})$, producing positive action value. This differential signal guides the actor's policy updates.

\subsubsection{Training Algorithm}

For the actor, our goal is to maximize the expectation of $q(s_t,a_t)$ for each step. So we use the clipped PPO loss~\cite{schulman2017proximal} with an entropy factor to encourage explorative actions:

{\small\begin{displaymath}
    L^\text{clip} = -\frac{1}{k}\sum_{t=0}^{k-1} \left[\min\left( \frac{\pi_\theta(a_t|s_t)}{\pi_{\theta_\text{old}}(a_t|s_t)}, \text{clip}\left(\frac{\pi_\theta(a_t|s_t)}{\pi_{\theta_\text{old}(a_t|s_t)}}, 1\pm\epsilon\right)\right) q_t \right]
\end{displaymath}}
\begin{displaymath}
    L^\text{entropy}=\frac{1}{k}\sum_{t=0}^{k-1}\pi_\theta(a_t|s_t)\log \pi_\theta(a_t|s_t)
\end{displaymath}
\begin{displaymath}
    L^\text{actor} = L^\text{clip}+\eta L^\text{entropy}
\end{displaymath}

For the critic, our goal is to minimize the bias of the evaluated state value $v_\phi(s)$ and the ground truth state value $v_\pi(s)$, so we use the MSE loss function to optimize the critic's parameters $w$:

\begin{displaymath}
    L^\text{critic}=\frac{1}{k}\sum_{i=0}^{k-1}(v_\phi(s_t)-v_\pi(s_{t}))^2
\end{displaymath}

We observe that the policy gradient for updating the actor depends on $q$ (calculated from the critic's state values), while the critic's update gradient relies on $v_\pi$ (determined by the actor's current policy). As shown in Algorithm~\ref{alg:ppo}, this bootstrapping interaction allows both components to progressively improve through repeated optimization iterations, where $\gamma$ is set to 1 for simplicity.

\begin{algorithm}[t]
    \caption{The training algorithm in LQRS}
    \label{alg:ppo} 
    \begin{algorithmic}[1]
        \Require Initial policy parameters $\theta$, initial value parameters $\phi$ 
        \Require Clip parameter $\epsilon$, training epochs $e$
        \State Collect episode $(s_0,a_0,r_1,s_1,\cdots,a_{k-1},r_k,s_k)$ using current policy $\pi_{\theta}$
        \State Get actual state values $V_\pi:(v_\pi(s_0),v_\pi(s_1),\cdots v_\pi(s_k))$, $ v_\pi(s_i)=\sum_{j=i+1}^k r_j+\sqrt{T_\text{execute}}$
        \State Get inferred state values $V_\phi:(v_\phi(s_0),v_\phi(s_1),\cdots v_\phi(s_k))$, $v_\phi(s_i)$ is inferred from critic net's initial parameters $\phi$
        \State Compute action values $Q = (r_1+v_\phi(s_1)-v_\phi(s_0),r_2+v_\phi(s_2)-v_\phi(s_1),\cdots ,r_k+v_\phi(s_k)-v_\phi(s_{k-1}),0)$
        \State Record old action probabilities $\pi_\text{old} = \pi_{\theta}(a_t|s_t)$ inferred from actor net's initial parameters $\theta$
        \For{epoch $=1$ to $e$} 
            \State Compute new probabilities $\pi_{\theta}(a_t|s_t)$
            \State Calculate probability ratio $r_t = \pi_{\theta}(a_t|s_t) / \pi_{\theta_\text{old}}(a_t|s_t)$
            \State Compute actor loss: $L^\text{actor}$
            \State Compute critic loss: $L^\text{critic}$
            \State Update $\theta \leftarrow \theta - \alpha \nabla_{\theta}L^\text{actor}$
            \State Update $\phi \leftarrow \phi - \beta \nabla_{\phi}L^\text{critic}$
        \EndFor
    \end{algorithmic}
\end{algorithm}

From the above analysis, we see that the critic evaluates \emph{each individual action} taken by the actor, providing feedback that is $k$ times denser than that of end-to-end reward methods. When the actor makes a suboptimal decision (e.g., selecting an inefficient join order), it receives immediate feedback via the critic's updated state value, without needing to wait for the entire query to complete. Which means even if subsequent actions compensate for earlier mistakes (e.g., correcting a bad join order later in the plan), the initial suboptimal action still receives its own critical feedback. This ensures that every action contributes to policy updates, not just the final outcome.

\subsection{Model Design}

This section presents the implementation details of the actor's policy $\pi_\theta(s)$ and the critic's value function $v_\phi(s)$.

When the decision model receives a partial plan $s$, it first encodes each of its nodes into embedding vectors. The resulting vectorized tree is then processed separately by the actor and critic networks. The actor generates a probability distribution $\pi_\theta(s)$ over the set of valid actions. Meanwhile, the critic produces a scalar estimate $v_\phi(s)$, representing the state's optimization potential.

\begin{figure}[t]
\centerline{\includegraphics[width=\linewidth]{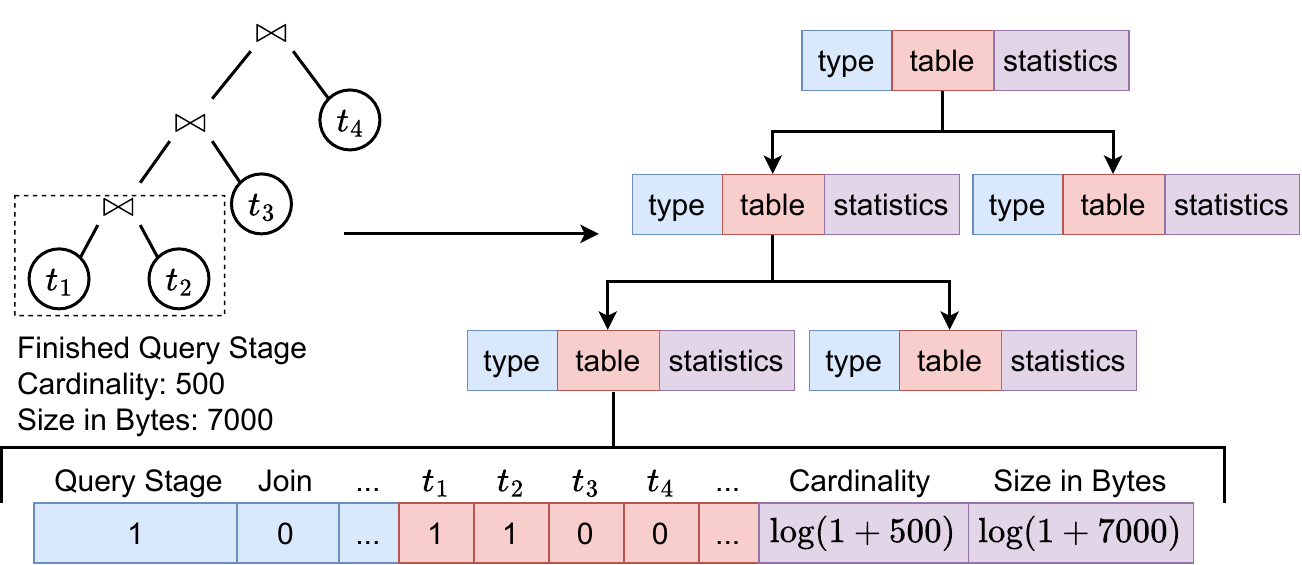}}
\caption{Feature encoding.}
\label{fig:featureencoding}
\end{figure}

\subsubsection{Feature Encoding}

The plan tree is transformed into a vector tree, as shown in Figure~\ref{fig:featureencoding}. Each node $u$ is encoded as a vector by concatenating three components:
\begin{displaymath}
    \text{encode}(u) = \text{type}(u) || \text{table}(u) || \text{statistics}(u)
\end{displaymath}
with the following encoding rules:

\begin{itemize}[nosep, leftmargin=*]
    \item \textbf{\textit{type$(u)$}} is a one-hot vector encoding the type of node $u$, such as a join or a query stage. In the simplified example shown in Figure~\ref{fig:featureencoding}, if $u$ is a leaf node, then type$(u)=(0,1)$; if $u$ is a join node, then type$(u)=(1,0)$.

    \item \textbf{\textit{table$(u)$}} is a binary (0/1) vector indicating which tables participate in node $u$, as shown in Figure~\ref{fig:featureencoding}. It encodes the join sequence leading to node $u$, and during AQE, even leaf nodes may touch multiple tables.

    \item \textbf{\textit{card$(u)$}} is set to $(\log(1+c))$ if $u$ corresponds to a finished query stage with observed cardinality $c$; otherwise, card$(u)$ is set to $(-1)$. The same encoding strategy is applied to other statistics, such as size in bytes.
\end{itemize}

This encoding strategy ensures that LQRS remains operational under data updates and schema changes. Specifically, the operator type $\text{type}(u)$ is fixed, while the intermediate cardinality $\text{card}(u)$ is collected online during execution rather than relying on precomputed statistics. Even when new tables are introduced, the encoding remains valid, with the corresponding positions taking a default value of 0.


The encoded vector tree is then processed by LQRS's actor and critic networks, both of which are instantiated using a TreeCNN architecture. 
We refer the reader to ~\cite{neo} for a deeper investigation into tree convolution applied to query plans.

\subsubsection{Masked Output}
\label{sec:curriculum}

To prevent invalid actions such as swapping nonexistent nodes, we apply masking to the actor's output. For a workload with at most $n$ tables, the dimension of the actor's output is given by:
\begin{displaymath}
    d_{\pi_\theta(s)}=2 + (n-1) + C_n^2 + n + 1
\end{displaymath}
for 2 \textit{init} actions, $n-1$ possible \textit{lead} actions(if we only consider lead a single table each time), $C_n^2$ possible \textit{swap} actions (each concrete action instance is treated as a distinct action, e.g., $\text{swap}(4,7)$), $n$ possible \textit{broadcast} actions, and 1 \textit{no-op} action. For example, when processing a partially executed plan with 3 leaf nodes, we create a binary mask as follows:

\begin{displaymath}
    \begin{split}
        M = (&\underbrace{0,0}_{\text{init(1), init(0), }},\underbrace{1,1,0,\cdots,1,0}_{\text{swap(1,2), swap(1,3), swap(2,3)}}, \\
             &\underbrace{1,1,0,\cdots}_{\text{lead(2), lead(3)}},\underbrace{1,1,1,0,\cdots}_{\text{broadcast(1), broadcast(2), broadcast(3)}}, \underbrace{1}_{\text{no-op}})
    \end{split}
\end{displaymath}

However, to achieve smoother policy convergence, we implemented a curriculum learning framework that requires masking out additional actions at different training stages. Initially, the model is only allowed to choose the plan initialization strategy, focusing on a simple binary decision task. During the mid-training phase, we progressively lift the masking on runtime plan adjustment actions; as these decisions are based on actual runtime statistics, they are generally easier for the model to learn. In the final training stage, all restrictions are removed except for masking invalid actions, allowing the model to explore the full action space and refine its decision-making capabilities. The final action probabilities are calculated by normalizing the masked output:

\begin{displaymath}
    \pi_\text{final}(s)=M\odot\pi_\theta(s)
\end{displaymath}

%% file: PlannerExtension.tex
\section{Planner Extension}

Our planner extension operates through two core mechanisms: extracting and sending the current partial plan to the decision model and applying optimization actions returned by the actor to modify the query plan.

The planner extension is built on top of Spark AQE. In particular, the planner extension can rely on AQE’s native support for pausing query execution at well-defined synchronization points and triggering re-optimization based on runtime statistics. In Spark AQE, a series of built-in optimization rules (including dynamic join selection) is applied in a pipelined fashion each time AQE is triggered. Our planner extension is integrated as an extension rule of Spark AQE to implement five optimization actions, without requiring the planner extension to explicitly manage execution suspension or execution state preservation.


\subsection{Send the Extracted Plan}

Upon triggering the planner extension rule, it first checks whether the maximum number of allowed optimization steps has been reached. If not, it extracts the current partially executed plan along with available runtime statistics from each completed query stage. The extracted plan and statistics are then sent to the decision model, after which the planner extension awaits further actions.

The setting of the maximum number of optimization steps plays a critical role in overall performance: higher values allow more optimization attempts and richer feedback signals, but they also require consistently accurate decisions across diverse scenarios. By default, we empirically set the maximum number of optimization steps (both before and during execution) to 3.

\subsection{Inject Plan}

While \textit{init} action and \textit{no-op} action are straightforward to handle, different strategies are required for the \textit{swap} action, \textit{lead} action, and the \textit{broadcast} action.

\subsubsection{Plan Transformation}
\label{sec:bushy}

The \textit{swap} action is denoted as swap$(i,j)$, indicating the exchange of positions between the $i$-th and $j$-th leaf nodes. However, if the opposite sides of the join of these two leaf nodes lack compatible join keys, executing the swap would result in a Cartesian product. The same issue also occurs while applying the \textit{lead} action. To mitigate this, we design Algorithm ~\ref{alg:bj} to perform controlled plan transformation.




\begin{algorithm}[t]
    \caption{The plan transformation algorithm in LQRS}
    \label{alg:bj} 
    \begin{algorithmic}[1]
        \Require Plan $p$ with leaf nodes $[l_1, l_2,\cdots,l_n]$ and join conditions $\{c_1,c_2,\dots,c_m\}$, swap action swap$(i,j)$ or lead action lead$(i)$
        \Ensure New plan $p'$ if the $i$-th and the $j$-th leaf nodes can be swapped or the $i$-th leaf node can be leaded, otherwise original plan $p$
        \State $L: [l_1, l_2,\cdots,l_n], C:\{c_1,c_2,\dots,c_m\} \leftarrow$ extractJoins($p$)
        \State
        $L'\leftarrow [l_1,\cdots,l_{i-1},l_j,l_{i+1},\cdots,l_{j-1},l_i,l_{j+1},\cdots,l_n]$ for swap$(i,j)$ or $L'\leftarrow [l_i, l_1,l_2\cdots,l_{i-1},l_{i+1},\cdots,l_n]$ for lead$(i)$
        \State Initialize $p' \leftarrow l_1$
        \For{$k=2$ to $n$} 
            \State Find condition $c \in C$ that cannot be evaluated separately on $p'$ or $L'_{k}$ but can be evaluated on $p'\bowtie L'_{k}$
            \If{$c$ exists}
                \State $p'\leftarrow$ buildJoin($p',L'_{k},c$)
            \Else
                \State \Return $p$
            \EndIf
        \EndFor
        \State \Return $p'$
    \end{algorithmic}
\end{algorithm}

We observe that applying \textit{swap} or \textit{lead} action during query execution can gradually transform the join plan into a bushy structure. As an illustrative example, consider the \textit{swap} action in Figure ~\ref{fig:ld2bushy}:
\begin{figure}[htbp]
    \centering
    \includegraphics[width=\linewidth]{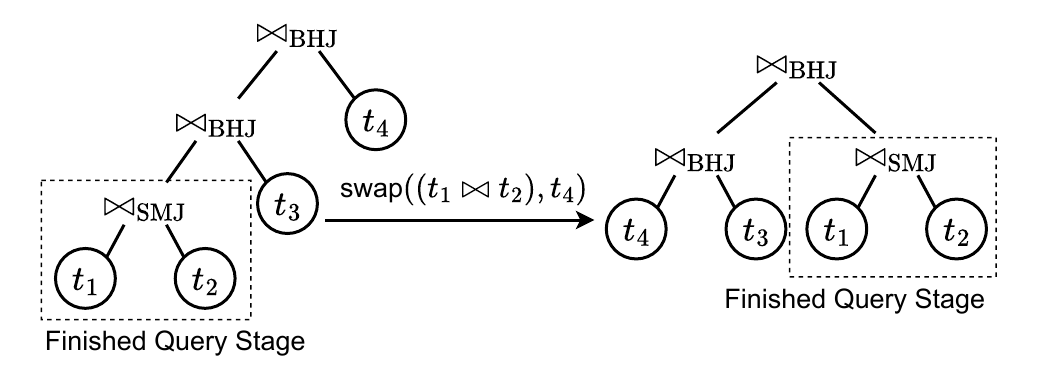}
    \caption{A swap between $(t_1\bowtie t_2)$ and $t_4$ can turn the left-deep join plan into a bushy join plan.}
    \label{fig:ld2bushy}
\end{figure}
By applying the \textit{swap} action between the completed $t_1 \bowtie t_2$ subtree and $t_4$, the planner extension incrementally constructs a bushy plan during execution. 

\subsubsection{Join Operator Selection}

Upon receiving a \textit{broadcast} action, the planner extension performs a bottom-up traversal of the query plan. It first locates the corresponding join operator within the execution plan hierarchy, then annotates the identified join operator with a \texttt{BROADCAST} hint on the appropriate (left/right) side. This modification instructs Spark SQL to prioritize the broadcast join strategy for the specified join operation.

\subsubsection{After Plan Injection}

After the optimization actions are applied, the modified plan is handed back to the AQE execution engine for continued execution. LQRS does not restart query execution or invalidate completed query stages. Instead, the injected modifications only affect the remaining unexecuted portion of the plan.

In addition, the planner extension benefits from AQE’s built-in plan validation mechanisms. While LQRS does not rely on Spark’s internal plan evaluation or cost comparison and always adopts the plan produced by its optimization actions, AQE still enforces a set of critical safety checks before execution is resumed. In particular, the AQE engine guarantees the semantic equivalence of the modified plan with respect to the original one and ensures that the injected plan can be correctly compiled and executed by the underlying engine. As a result, the planner extension does not need to handle low-level concerns such as semantic equivalence checking, plan code generation, or execution resumption. 

\subsection{Compatibility with Other Systems}

While the planner extension in LQRS is implemented on top of the Spark AQE framework, its design principles are applicable to other database systems. A key challenge in generalizing such extensions lies in identifying points at which execution can be paused or re-optimized with acceptable overhead. Spark SQL naturally provides such boundaries through query stages, which explicitly materialize intermediate results and support execution suspension and re-optimization. For database systems such as Flink SQL in batch mode, Presto, and Snowflake, which have similar execution models, natural synchronization points arise where intermediate results are produced. In these systems, planner extensions can likewise use these materialization boundaries to perform runtime optimization.

For other systems that do not provide such exploitable execution boundaries, we need to identify points where execution can be safely paused or re-optimized. In PostgreSQL, We observe that pipeline-breaking operators, such as hash and sort operators, can be synchronization points because they inherently materialize intermediate results and interrupt tuple pipelines. DuckDB employs a vectorized execution engine, yet certain operators still require blocking behavior, creating implicit plan boundaries where runtime statistics can be observed. These observations indicate that even without explicit AQE support, our planner extension can use existing materialization or pipeline breakers for re-optimization.

%% file: Experiments.tex
\section{Experiments}

\subsection{Experiment Setup}

\subsubsection{Environment}

All experiments were conducted on Spark 3.5.4 with 6 executors, each configured with 6 cores and 20 GB memory. The decision model was trained and executed for inference on an NVIDIA A40 GPU.

\subsubsection{Benchmarks}

We selected four standard benchmarks, following prior work ~\cite{glo,hybrid, chen2024lero}:

\noindent\textbf{JOB}~\cite{10.14778/2850583.2850594}: JOB employs the real-world IMDb dataset with 21 tables related to movies and actors. It includes 113 complex SQL queries generated from 33 templates, each of them involving joins across 4 to 17 tables.

\noindent\textbf{Extended JOB} (referred to as \textbf{ExtJOB})~\cite{neo}: ExtJOB is an extension of the JOB benchmark, consisting of 24 queries derived from 12 distinct templates. Compared to the original JOB queries, ExtJOB features entirely different join graphs and predicates, making it a more challenging benchmark~\cite{balsa}.

\noindent\textbf{STACK}~\cite{marcus2021bao}: STACK comprises 10 tables from Stack Exchange datasets. Its 16 query templates generate analytical queries joining 4-12 relations. We excluded templates \#4,7 (with arithmetic operations in join conditions) and \#9,10 (containing unsupported predicate subqueries)—limitations shared by most existing LQOs. We plan to address these restrictions in the future.

\noindent\textbf{TPC-H}~\cite{tpc-h}: TPC-H is designed to test the response time of database systems to complex queries. It includes eight tables with the scale factor set to 100 in our experiments. In our evaluation, we only consider the query templates \#2, 3, 5, 7, 8, 9, 10 following~~\cite{chen2024lero}. We exclude other templates, as they are too simple (only have one or two tables) or contain predicate subqueries.

\subsubsection{Baselines}

We compare LQRS against four baselines:

\noindent\textbf{Spark SQL}: Spark SQL with its default AQE mechanism.

\noindent\textbf{Lero}~\cite{chen2024lero}: Lero employs a learned comparator to evaluate join plans generated by Spark SQL's optimizer under different cardinality estimations, then chooses the best one among them.

\noindent\textbf{AutoSteer}~\cite{anneser2023autosteer}: AutoSteer systematically evaluates all available optimization rules in Spark SQL by disabling them to assess their impact on the current plan. It then constructs a collection of rules to disable for performance gains using greedy search.

\noindent\textbf{SSA}: QuerySplit~\cite{querysplit} is a query re-optimization algorithm that proactively decomposes a query into multiple subqueries at the logical plan level and iteratively selects subqueries to execute based on their estimated cost and output cardinality. In Spark SQL, query plans is naturally partitioned into query stages at shuffle boundaries. Therefore, we only re-implement the subquery selection component of QuerySplit, namely SSA, within Spark SQL and use it as a query re-optimization baseline for comparison.


\subsubsection{Settings}
For static evaluation, we refer to the experimental settings of Bao~\cite{marcus2021bao} and Lero~\cite{chen2024lero}, while for dynamic evaluation, we refer to the settings used in Robust-MSCN~\cite{robust}:

\noindent\textbf{Data Generation.} For JOB and ExtJOB, we expanded the original IMDb dataset to ten times its original size. For STACK, we utilized the benchmark's original archived dataset. For TPC-H, we generated the dataset using the official data generation tool with a scale factor of 100. All datasets were converted to Parquet format. Note that this experimental setup already imposes substantial memory pressure on our testing environment, as a bad query plan can easily lead to out-of-memory errors and execution failure.

\noindent\textbf{Query Generation.} We generated 1,000 training queries using each benchmark's templates. For each template, randomized predicate conditions were introduced while preserving the join structure. For JOB and ExtJOB, we exclusively used their original 113 and 24 queries as their test set, respectively. For STACK and TPC-H, the test set contains 10 additional queries per template generated in the same manner. 

\noindent\textbf{Evaluation Approach.} We conduct both static and dynamic evaluations. For the static evaluation, each optimizer is trained on the training set of a given benchmark until convergence, and then its performance is assessed by executing the corresponding test queries. For the dynamic evaluation, we train the optimizer on one dataset and then test it either on the test set of another dataset or on the same dataset after dynamically modifying the underlying data through updates.

\noindent\textbf{Key Metrics.} We evaluate performance using \emph{End-to-end execution time:} total time required to run all test queries. Each query’s end-to-end execution time is capped at 300s; queries that exceed this limit or fail due to OOM error are recorded as 300s.

\subsection{Query Performance}

Table~\ref{tab:overall} reports the end-to-end execution time across all benchmarks.
\begin{table}[t]
  \caption{Overall performance on all benchmarks.}
  \label{tab:overall}
  \begin{tabular}{ccccc}
    \toprule
    Time (s)& JOB & ExtJOB & STACK & TPC-H\\
    \midrule
    Spark SQL & 5821.4 & 694.0 & 13149.1 & 2954.0\\
    Lero & 33211.9 & 752.9 & 4231.6 & 2625.0\\
    AutoSteer & 7793.3 & 903.2 & 13026.1 & 3010.2\\
    SSA & 3124.1 & 651.0 & 7934.2 & 4895.7 \\
    LQRS & \textbf{3069.5} & \textbf{555.5} & \textbf{3504.3} & \textbf{2471.3}\\
    \bottomrule
  \end{tabular}
\end{table}
LQRS consistently achieves the best performance on all four workloads, outperforming all baseline methods. Compared with Spark SQL, LQRS reduces the total execution time by 16.3\%–73.3\% across four benchmarks. These results highlight the advantage of LQRS’s novel
idea of allowing learned optimization decisions to be deferred to execution time and guided by actual runtime observations.

Figure~\ref{fig:grid} illustrates the performance improvement and regression of each query on four tested benchmarks with Lero, AutoSteer, SSA, and LQRS w.r.t the corresponding end-to-end query execution latency in Spark SQL. We omit cases where Spark SQL chooses a similar plan (with a runtime difference $<$ 5s) to LQRS and other baseline methods. The key observations are as follows:

\begin{figure*}[t]
    \centering
    \includegraphics[width=\linewidth]{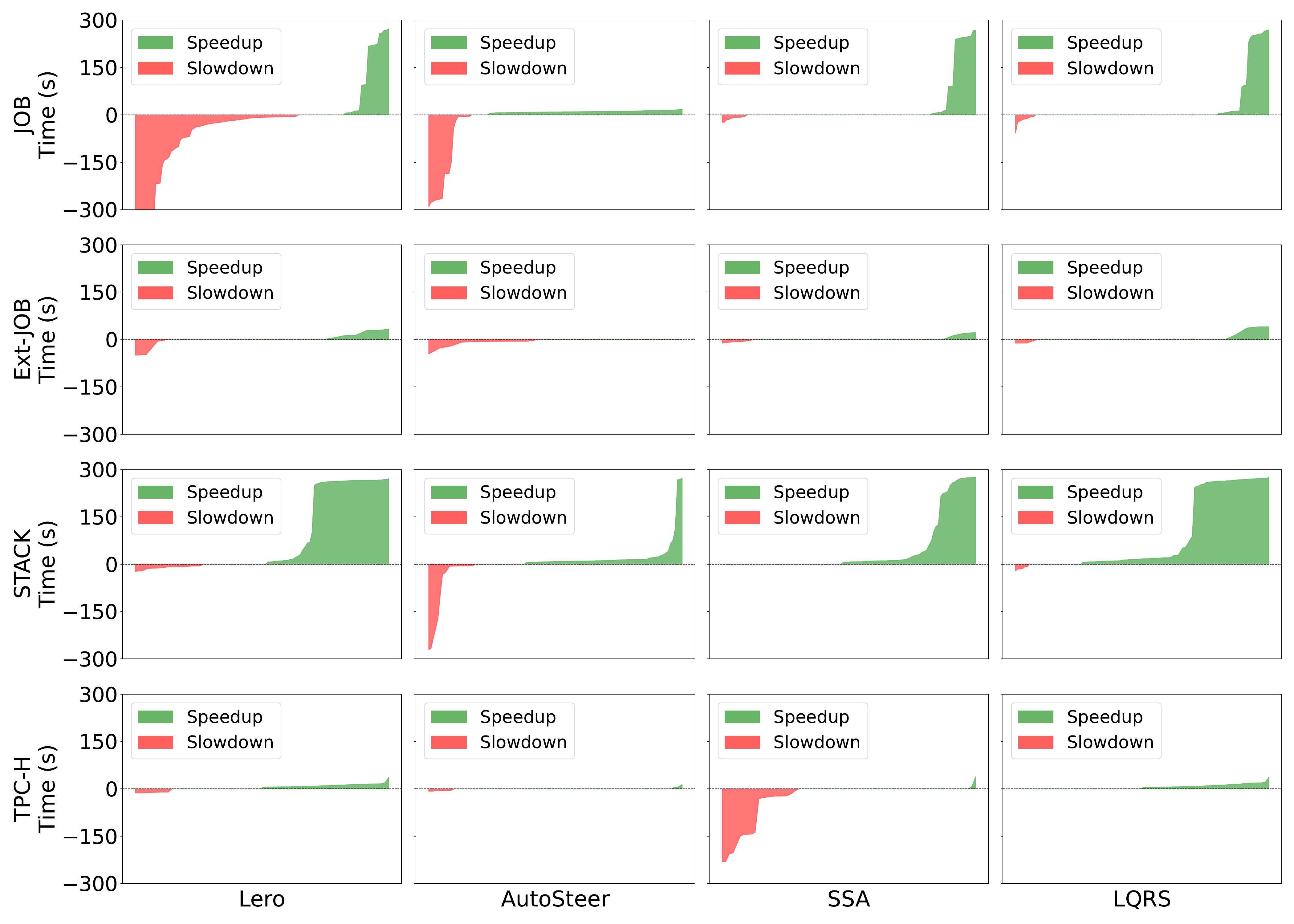}
    \caption{Latency improvement and regression of each query in Lero, AutoSteer, SSA, and LQRS w.r.t Spark SQL}
    \label{fig:grid}
\end{figure*}

\begin{itemize}[nosep, leftmargin=*]
    \item Compared with Lero, our performance improvements exhibit similar trends, but we observe fewer regressions, especially on JOB. This indicates that LQRS can leverage runtime observations to obtain execution plans of comparable quality to Lero, while avoiding the extensive plan enumeration and evaluation overhead that Lero incurs, highlighting the effectiveness of LQRS.
    \item Compared with AutoSteer, our improvements are substantially larger. AutoSteer's learned optimization strategy tends to favor disabling high-overhead rules, which often backfires on complex queries. In contrast, LQRS does not generate any inferior plans during test query execution, highlighting the robustness of its optimization strategy.
    \item Compared with SSA, our performance gains are more substantial and stable, with fewer large-scale regressions observed on TPC-H. This is because our reinforcement learning–based decision process is more comprehensive: it evaluates feedback at each step of the actor and considers the long-term value of partially executed plans. As a result, LQRS does not greedily lead low-cost leaf nodes forward, but instead balances immediate costs with long-term benefits, highlighting the intelligence of its decision-making process.
\end{itemize}

\subsection{Case Study}

In this section, we analyze the results on a per-query basis to find out what LQRS actually achieves in optimizing query plans. 

\subsubsection{Query Improvement Analysis}
\label{sec:indirect}

\begin{figure}[t]
    \centering
    \includegraphics[width=\linewidth]{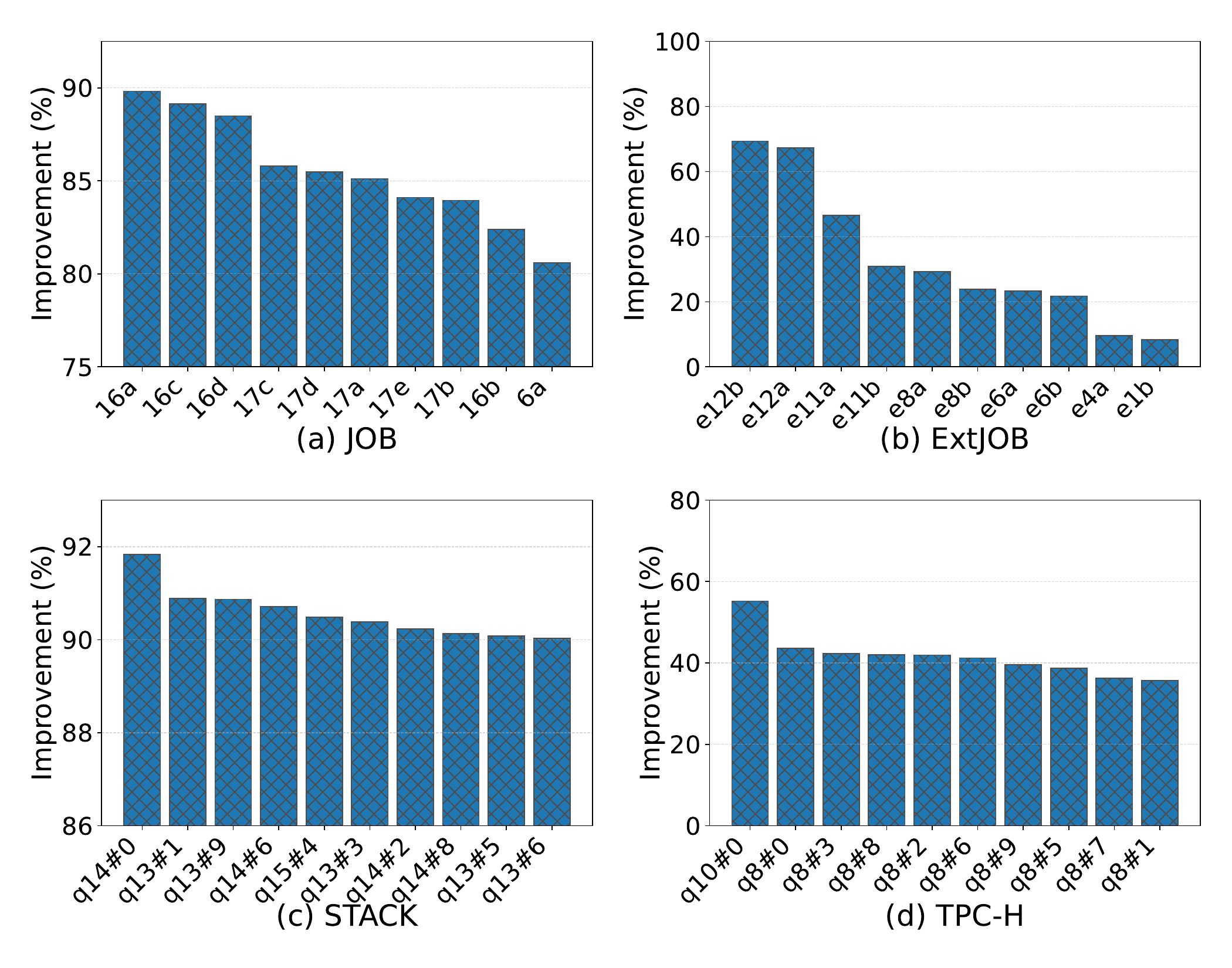}
    \caption{The top 10 queries improved by LQRS in each benchmark compared with Spark SQL.}
    \label{fig:top10}
\end{figure}

Figure ~\ref{fig:top10} shows the top 10 queries improved by LQRS in each benchmark compared with Spark SQL. LQRS achieves up to 89.81\%, 69.33\%, 91.83\%, and 55.21\% performance improvement on the JOB, ExtJOB, STACK, and TPC-H benchmarks, respectively.

For example, in JOB query Q16a, LQRS not only chooses the cost-based initialization strategy but also refines the plan by prioritizing the join between table \textbf{\texttt{t}} and table \textbf{\texttt{mk}}, significantly reducing the execution time from over 300s to just 30.56s, demonstrating LQRS’s ability to refine CBO-generated plans by improving join orders while balancing optimization overhead and performance gains.

LQRS’s improvements are not limited to join ordering, even in the absence of the \textit{broadcast} action . When smaller intermediate results are produced earlier in the plan, AQE is more likely to convert subsequent SMJs into more efficient BHJs. As shown in the STACK query q1\#1, LQRS transforms the original join order \textbf{\texttt{t}} $\bowtie$ \textbf{\texttt{s}} $\bowtie$ \textbf{\texttt{q}} $\bowtie$ \textbf{\texttt{tq}} into \textbf{\texttt{t}} $\bowtie$ \textbf{\texttt{tq}} $\bowtie$ \textbf{\texttt{q}} $\bowtie$ \textbf{\texttt{s}} through two lead(3) actions and one lead(2) action. This not only produces a better join order but also changes the original two SMJs and one BHJ into one SMJ and two BHJs, effectively improving the query’s execution time by 45.03\%. The reason is that in the latter join order, one side of the third join has only 4 MB of data, whereas in the former, the two sides contain 281.1 MB and 1125.6 MB, respectively. As a result, AQE dynamically switches this join from an SMJ to a BHJ during execution. Similar phenomena can be observed in many other queries. For instance, in STACK query q13\#1, LQRS lead the \textbf{\texttt{a}} table to the front of the join sequence, transforming the original plan with only one BHJ into a new plan with four BHJs. This new join order and strategy combination reduced the execution time from over 300s to just 41.94s. Similar improvements can also be seen in queries such as ExtJOB query Qe11b, JOB query 6a etc. These results demonstrate that LQRS can consecutively refine join orders during execution and cooperate effectively with native AQE optimizations to further improve join strategies.

\subsubsection{Query Regression Analysis}

While LQRS delivers substantial improvements on many queries, we also observe performance regressions on a small subset. Minor regressions are mainly due to the optimization overhead introduced by LQRS.

Significant regressions in query performance mostly stem from inherent limitations of LQRS’s current learning framework, which occasionally results in suboptimal plan selections. 
For instance, JOB Q8d experiences performance degradation because the table \textbf{\texttt{t}} was incorrectly selected as the lead table, causing a substantial increase in intermediate data movement. As a result, the shuffle size during query execution nearly tripled compared to the baseline plan, significantly impacting runtime.

To mitigate such regressions, we introduced several training-time constraints, including limiting the number of candidate lead tables and capping the maximum number of action steps. These restrictions guide the agent toward more stable and predictable decisions while reducing exploration noise. Nevertheless, regressions of this kind cannot be entirely avoided under the current framework and represent an important avenue for future improvement.

\subsubsection{Proportion of bushy plans}

Here, we analyze the proportion of test queries whose final executed plans adopt a bushy join structure. In our experiments, 22 (19.5\%) queries in JOB, 5 (20.1\%) in ExtJOB, and 30 (27.3\%) in STACK ultimately execute with bushy plans, enabled by the plan transformation actions as described in Section~\ref{sec:bushy}. These results show that LQRS is capable of effectively exploring bushy join plans to improve query performance. 

\subsection{Dynamic Evaluation}

\begin{figure}[t]
    \centering
    \includegraphics[width=\linewidth]{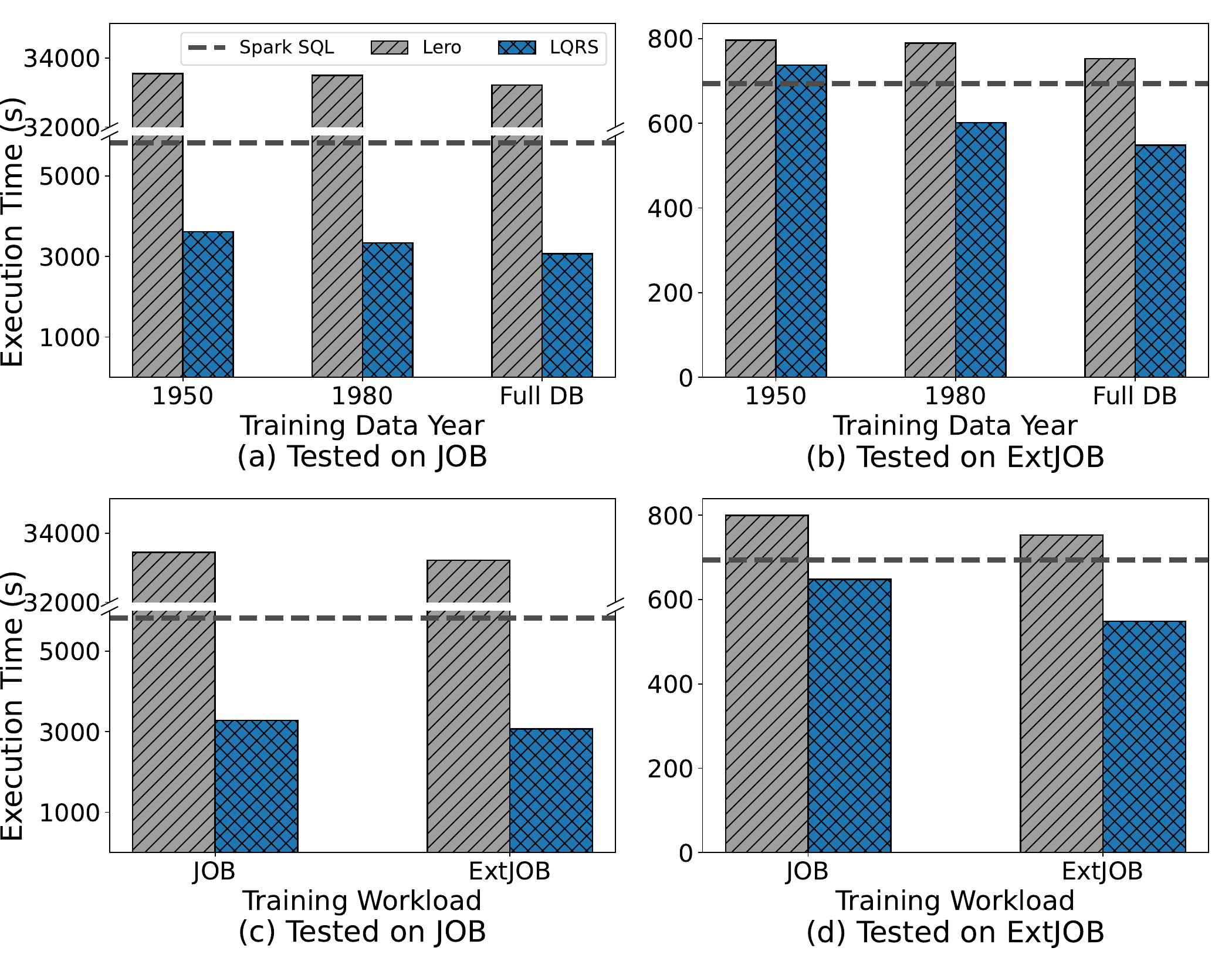}
    \caption{Performance of LQRS and Lero under data distribution and workload changes. (a–b) Optimizers trained on older IMDb snapshots (1950, 1980) and tested on the full database. (c–d) Optimizers trained on either JOB or ExtJOB workloads and tested on both JOB and ExtJOB.}
    \label{fig:dynamic}
\end{figure}

Figure~\ref{fig:dynamic} evaluates the robustness of LQRS and Lero under data distribution and workload changes. We chose Lero due to its relatively strong adaptability among all the baselines.

For data distribution changes (Figure~\ref{fig:dynamic}(a–b)), we construct two historical variants of IMDb—IMDb-1950 and IMDb-1980—by retaining only movies released no later than 1950 and 1980, respectively, together with their related records. These variants contain less than 10\% and about 30\% of the full IMDb dataset~\cite{robust}. The optimizers are trained on these historical datasets and evaluated on the full IMDb to assess robustness to distribution shifts. For workload changes (Figure~\ref{fig:dynamic}(c–d)), the optimizers are trained on either JOB or ExtJOB and evaluated on both workloads, enabling cross-workload generalization analysis.

We observe that LQRS consistently outperforms both Spark SQL and Lero under various experimental settings, except for a slight performance drop below Spark SQL when trained on ExtJOB queries using the IMDb-1950 dataset and tested on the full IMDb database. This result demonstrates LQRS’s robustness to environmental changes. Such robustness mainly stems from two factors: (1) the use of runtime observation enables LQRS to adapt its optimization behavior to the current execution context; and (2) the simplified feature encoding architecture effectively prevents overfitting.

However, both learning-based approaches suffer performance degradation under execution environment changes, with LQRS exhibiting relatively larger drops in raw execution performance. Across the four settings, Lero’s maximum degradation ranges from 7.53\% to 10.80\%, whereas LQRS experiences more severe drops of up to 34.93\%. In many cases, LQRS fails to select the optimal plan because the required action was rarely explored during training, whereas Lero can almost always ensure that even an unchosen plan at training at least belongs to its explored set. If LQRS were trained using an offline RL algorithm equipped with a comprehensive experience buffer, this limitation might be mitigated.

\subsection{Ablation Study}

In this ablation study, we will try to replace LQRS's core components with other possible substitutes to test out the effectiveness of LQRS's design on ExtJOB. We choose ExtJOB because it is relatively small, making experiments more convenient to run, yet it is still sufficient to reveal performance trends. LQRS uses PPO as the default reinforcement learning algorithm, a TreeCNN architecture for both actor and critic and action networks, and an action space that includes \textit{init} action, \textit{lead} action (with only one allowed leaf node), and \textit{no-op} action. Learning is conducted through the curriculum process described in Section~\ref{sec:curriculum}.

\subsubsection{Different Reinforcement Learning Algorithms}

\begin{figure}[t]
    \centering
    \includegraphics[width=\linewidth]{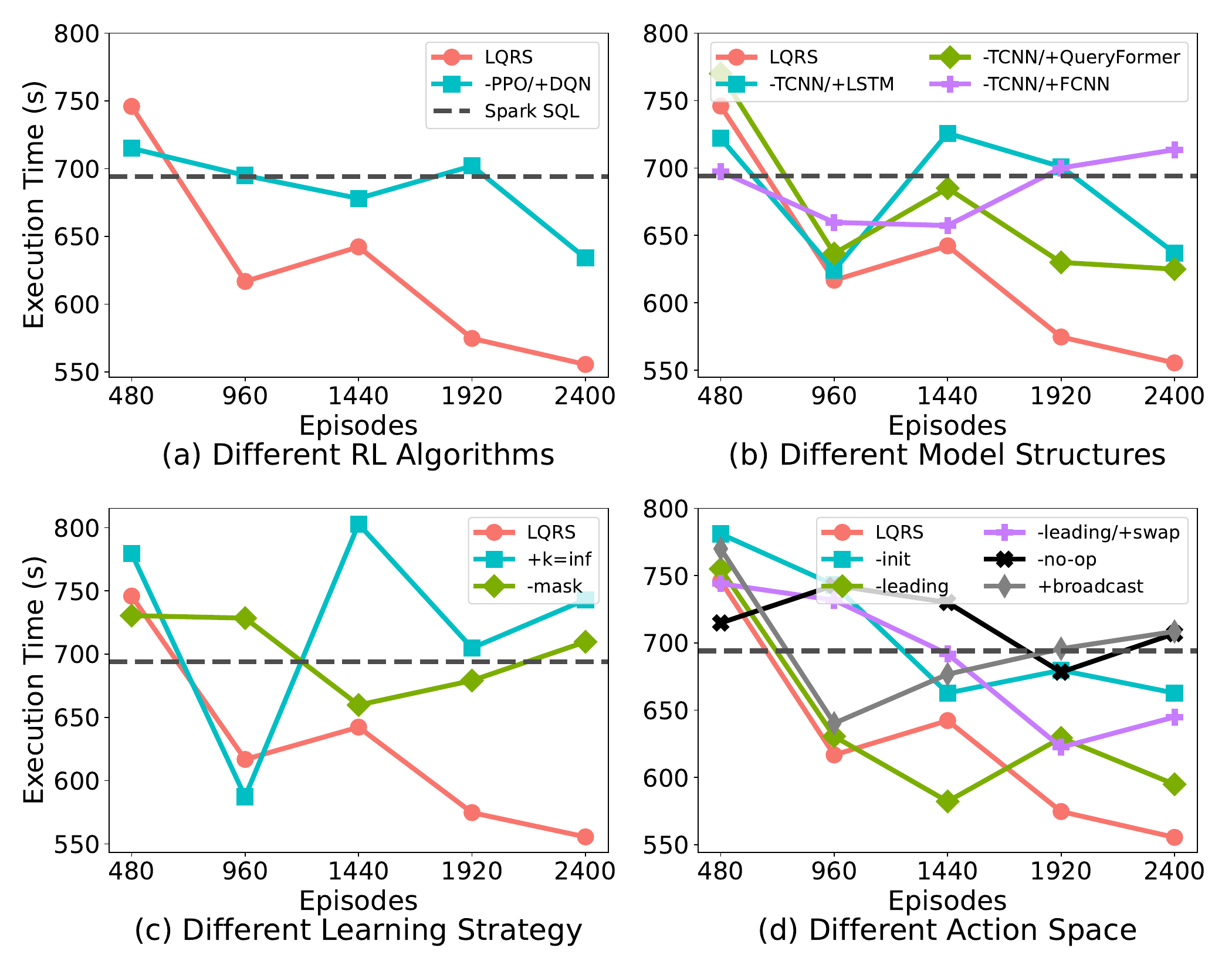}
    \caption{Query performance on ExtJOB with different substituted modules, under varying numbers of training queries.}
    \label{fig:ablation}
\end{figure}

To evaluate the impact of different reinforcement learning algorithms on query optimization performance, we compare the PPO algorithm used in LQRS with DQN ~\cite{mnih2015human}. As shown in Figure~\ref{fig:ablation}(a), LQRS with PPO surpasses Spark SQL in execution latency significantly earlier during training compared to its DQN-based variant. Also, after training on 2400 queries, the PPO-based model achieves consistently lower execution latency than the DQN-based model. Although DQN benefits from experience replay, it suffers from limitations in environments with large action spaces and non-stationary dynamics.

\subsubsection{Different Neural Network Structures}
\label{sec:model}

We experimented with several model architectures proposed in prior work, including a fully connected neural network (FCNN), TreeCNN ~\cite{marcus2021bao}, LSTM ~\cite{lstm}, and QueryFormer ~\cite{zhao2022queryformer}. To adapt these models for LQRS, we replaced their original feature encoding with our own and substituted their scalar output layers with a LogSoftmax layer. Among them, FCNN is the only model that fails to outperform Spark SQL even after training on 2400 queries. Specifically, LSTM exhibits the lowest optimization overhead, making it more efficient during both training and inference. However, TreeCNN demonstrates better convergence behavior and achieves higher overall performance after training, albeit with a slightly increased overhead.

\subsubsection{Different Learning Strategy}

To evaluate the effectiveness of our learning strategy design, we conducted an ablation study using two alternative strategies (Figure~\ref{fig:ablation}(c)). In the first strategy, we removed the step limit during optimization while retaining the curriculum learning schedule. In the second strategy, we limited the maximum number of steps but removed the curriculum constraint, allowing LQRS to explore the full action space from the beginning. 

Experimental results show that the first strategy performs comparably to the default configuration during the early training phase. However, as the curriculum progresses to its second stage, the model tends to take significantly more steps, making it difficult to repeatedly make correct decisions. The second strategy, despite enforcing a strict step limit, ultimately fails to outperform Spark SQL. This is because, in most cases, only a few targeted corrections are sufficient to reach an efficient plan. Without early-stage guidance, the model explores an overly large action space too freely, which hinders convergence.

\subsubsection{Different Action Space}

LQRS’s planner extension supports a wide range of action types. We conducted ablation studies to identify which subsets of the action space contribute most to learning efficiency and convergence. As shown in the results, enabling the \textit{broadcast} action often leads the model to generate plans that broadcast excessively large tables during training, resulting in significant performance fluctuations and instability. When the \textit{lead} action is disabled, LQRS loses its ability to explicitly manipulate join ordering, thereby reducing its optimization power and leading to suboptimal plans for many complex queries. Finally, disabling the \textit{init} action prevents LQRS from jumping to an entirely different planning path starting from a CBO-generated plan. This restriction increases the difficulty for LQRS to escape local optima and discover better execution plans.

\subsection{Overhead Study}

\begin{table}[t]
  \caption{Optimization cost on all benchmarks.}
  \label{tab:overhead}
  \begin{tabular}{ccccc}
    \toprule
    Time (s)& JOB & ExtJOB & STACK & TPC-H\\
    \midrule
    Lero & 30098.1 & 201.3 & 1110.6& 379.5 \\
    AutoSteer & 829.2 & 176.7 & 634.0 & 442.1\\
    SSA & 81.5 & \textbf{6.2} & 41.0 & 27.7\\
    LQRS & \textbf{33.3} & 7.1 & \textbf{37.3} & \textbf{19.6}\\
    \bottomrule
  \end{tabular}
\end{table}

As shown in Table~\ref{tab:overhead}, LQRS achieves the lowest optimization overhead among most evaluated methods, by making re-optimization decisions during query execution based on accurate runtime information. Its design explicitly supports a \textit{no-op} action, allowing the optimizer to skip unnecessary plan modifications when the expected benefit is limited, which effectively suppresses redundant re-optimizations and keeps the optimization process lightweight.

In comparison, SSA incurs higher overhead because it continuously modifies the plan whenever the current leading leaf node is not estimated as the lowest-cost node, resulting in unnecessary plan modifications during execution. Learned query optimizers such as Lero and AutoSteer suffer from even greater costs due to explicit enumeration and evaluation of a large plan search space before execution. Lero perturbs cardinality estimates at multiple levels to generate numerous alternative plans, while AutoSteer explores different combinations of optimization rules, both requiring repeated cost estimation over multiple candidates.

%% file: Conclusion.tex
\section{Conclusion}

We propose \emph{LQRS}, a learned query re-optimizer integrated into Spark SQL. LQRS serves as a prototype system that demonstrates how execution-time feedback and intervention can be incorporated into a modern query engine via a planner extension. By continuously observing partial execution results and applying query stage-level feedback with runtime-aware rewards, LQRS incrementally refines the remaining execution plan during query execution. Extensive experiments on Spark SQL demonstrate that LQRS reduces end-to-end execution time by up to 90\% compared to other learned query optimizers and query re-optimization methods.